\documentclass[%
 reprint,
 amsmath,amssymb,
 aps,
 pre,
]{revtex4-1}

\usepackage{graphicx}
\usepackage{dcolumn}
\usepackage{bm}

\usepackage[dvipsnames]{xcolor}

\begin{document}

\preprint{APS/123-QED}

\title{A Map Equation with Metadata:\\Varying the Role of Attributes in Community Detection}

\author{Scott Emmons}
 \email{scott3@live.unc.edu}
\author{Peter J. Mucha}%
 \email{mucha@unc.edu}
\affiliation{%
 Carolina Center for Interdisciplinary Applied Mathematics\\
 Department of Mathematics\\
 University of North Carolina\\
 Chapel Hill, NC 27599, USA
}%

\begin{abstract}
Much of the community detection literature studies structural communities, communities defined solely by the connectivity patterns of the network. Often, networks contain additional metadata which can inform community detection such as the grade and gender of students in a high school social network. In this work, we introduce a tuning parameter to the content map equation that allows users of the Infomap community detection algorithm to control the metadata's relative importance for identifying network structure. On synthetic networks, we show that our algorithm can overcome the structural detectability limit when the metadata is well-aligned with community structure. On real-world networks, we show how our algorithm can achieve greater mutual information with the metadata at a cost in the traditional map equation. Our tuning parameter, like the focusing knob of a microscope, allows users to ``zoom in'' and ``zoom out'' on communities with varying levels of focus on the metadata.
\end{abstract}

\maketitle

\section{Introduction}

As network science has found application in a variety of real-world systems, ranging from the biological to the technological, so too has community detection in networks received widespread attention \cite{porter_2009,fortunato_community_2010,fortunato_community_2016,shai_2017}. Traditionally, community detection methods have focused solely on the topology of the network, optimizing an objective function defined on the network structure that captures a particular notion of community, such as intra-community edge density and inter-community edge sparsity. Many approaches, ranging from the statistical to the information theoretical, have been used for community detection, and tradeoffs between these approaches include describing extant links versus predicting missing links \cite{ghasemian_evaluating_2018}.

More recent community detection work utilizes node metadata such as the grade and gender of students in a high school social network. As the No Free Lunch theorem states, community detection algorithms must make tradeoffs \cite{peel_ground_2017}, and node metadata can be used to guide community detection. For example, Newman and Clauset demonstrated that their stochastic block model (SBM) approach can choose either to partition a middle school and high school social network into communities by grade or into communities by race, depending on the metadata of interest \cite{newman_structure_2016}.
Similarly, Hric \textit{et al.} \cite{hric_network_2016} developed an attributed SBM from a multilayer perspective, with the attribute layer modeling relational information between attributes. Stanley \textit{et al.} \cite{stanley_2018} considered a different graphical model relating connections and attributes, with assumptions on the attribute distributions, to develop a stochastic block model with multiple continuous attributes.
Introducing the I-louvain method \cite{ilouvain}, Blondel \textit{et al.} extended the well-known Louvain algorithm \cite{blondel} for modularity maximization by including attributes in their ``intertia-based modularity.'' Yang \textit{et al.} proposed CESNA \cite{yang_community_2013} and He \textit{et al.} proposed CNMMA \cite{8609622} to identify communities by learning a latent space that generates links and attributes. Peel \textit{et al}. \cite{peel_ground_2017} established a statistical test to determine if attributes correlate with community structure, and they developed an SBM with flexibility in how strongly to couple attributes and community labels in the corresponding stochastic block model inference.
In related work, Stanley \textit{et al.} \cite{stanley_align_2018} propose a test statistic based on label propagation for the alignment of node attributes with connectivity patterns.

Prior work extending SBMs, such as the method of Newman and Clauset \cite{newman_structure_2016} and the method of Stanley \textit{et al.} \cite{stanley_2018}, are based on statistical relationships between metadata and network structure. Using such methods, a social science researcher who cares particularly about gender groups in a high school social network, for example, has no way to communicate a special interest in community alignment with the gender metadata. The methods will use the gender metadata insofar as it explains network structure, but otherwise the methods might ignore this feature. A key motivation for our work is to provide direct control over how much network communities align with a particular metadata type. With our method, the example researcher above can directly tune how much more she weights communities aligned with gender to communities describing network structure.

Ghasemian \textit{et al.} \cite{ghasemian_evaluating_2018} characterize this tradeoff between a statistical model of network formation, given by the algorithm of Newman and Clauset \cite{newman_structure_2016}, and an information-theoretic description of observed structure, given by the map equation \cite{rosvall_maps_2008}, as a tradeoff between under- and overfitting in community detection.  From the point of view of this framework, our method's contribution is enabling users to choose how much to overfit the metadata in describing the observed network structure.

Most closely related to our approach is the content map equation proposed by Smith \textit{et al.} \cite{smith_partitioning_2016}. The content map equation, as we later describe in more detail, adds an additional term to the map equation \cite{rosvall_maps_2008} that introduces entropy based on the metadata. This modification to the map equation encourages intra-module homogeneity of node metadata values.

Our paper extends the content map equation, categorizing the different sources of entropy in the map equation into the ``inter-module codebook,'' ``intra-module codebooks,'' and ``metadata codebooks.'' Within this framework, we introduce a tuning parameter to the metadata codebooks that allows direct control over the relative importance of particular metadata types. Similar to how focusing knobs are an essential feature of a microscope, adding a tuning parameter to the content map equation is essential to its function, allowing one to ``zoom in'' and ``zoom out'' on communities with varying levels of focus on the metadata.

\section{Methodology}

The map equation frames the problem of community detection as minimizing the description length of a random walk on the network \cite{rosvall_maps_2008}. In developing a code to compress the description of the random walk, the map equation necessitates that each codeword corresponds to an identifiable entity in the graph. It designates codewords for hard partitions of nodes into modules, codewords for individual nodes within each module, and codewords for a special ``exit'' keyword for each module. As the codeword for a given node needs only to be unique within that node's module, the module names and node names function like geographic city names and street names. The output of the map equation is a sort of ``map,'' optimized for data compression, that captures patterns in the data.

The map equation's entropy arises from two different types of codebooks. The ``inter-module codebook,'' consisting of module names, describes movement between modules. The ``intra-module codebooks,'' consisting of node names and special ``exit'' codewords, describe movement within modules. The sum of the entropies of these codebooks, weighted by their relative frequencies, gives the per-step average number of bits needed to describe an infinite random walk on the network for a given partition M of the nodes into $m$ modules:
$$L(\text{M}) = q_{\curvearrowright} H(\mathcal{Q}) + \sum\limits_{i=1}^m p_{\circlearrowright}^i H(\mathcal{P}^i)\,,$$
where $H(\mathcal{Q})$ is the entropy of the inter-module codebook, used with relative frequency $q_{\curvearrowright}$, and $H(\mathcal{P}^i)$ is the entropy of the intra-module codebook for module $i$, used with relative frequency $p_{\circlearrowright}^i$.

\begin{figure*}
  \includegraphics[width=0.71\linewidth]{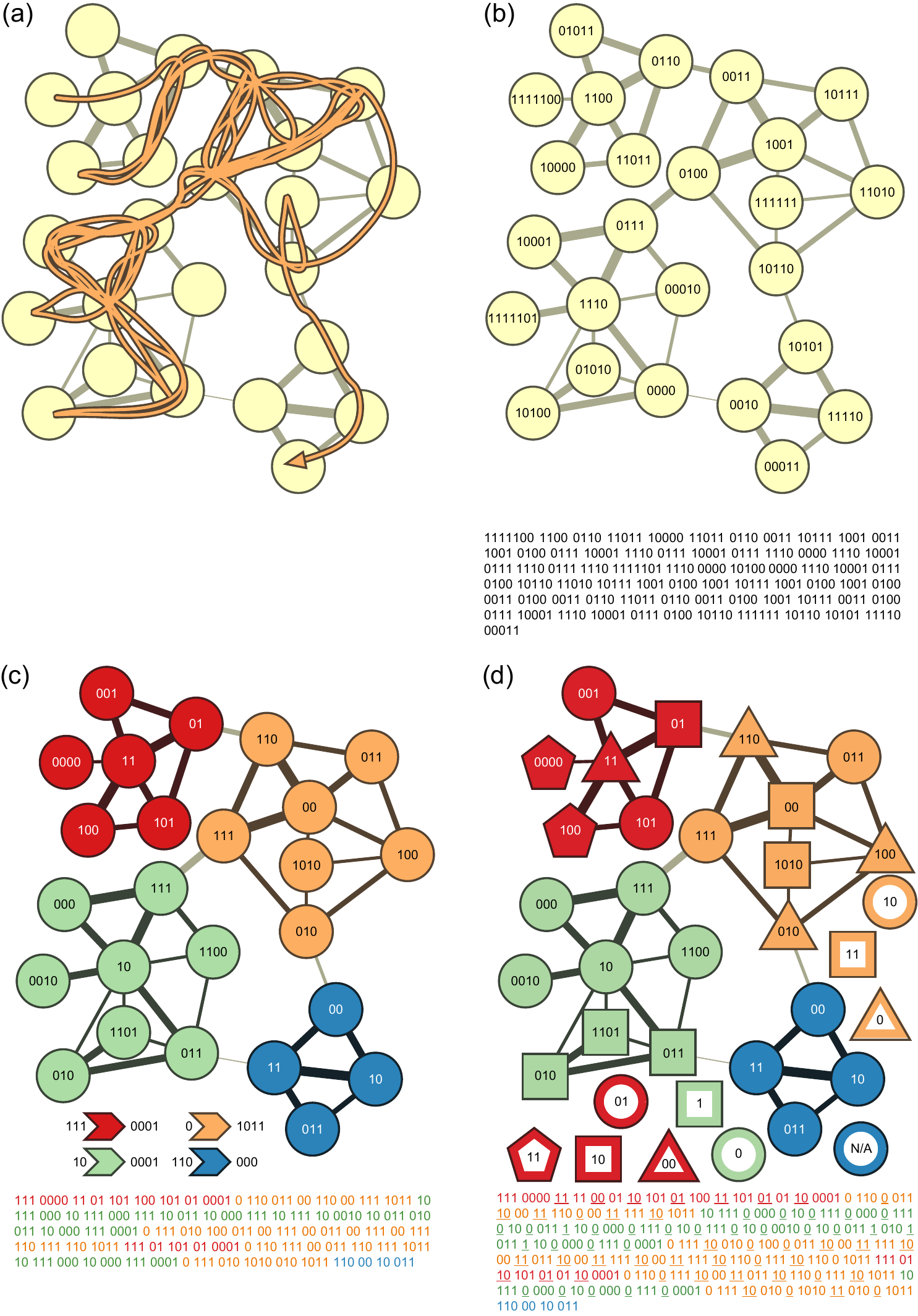}
  \caption{(Color online) Illustrating how our formulation of the content map equation extends the traditional map equation. (a) An example random walk on a weighted, undirected network. The thickness of each edge corresponds to its weight. (b) Encoding the location of the random walker with a Huffman encoding that assigns each node a unique codeword in a single codebook. (c) Compressing the description of the random walk with the traditional map equation, which encodes the location of the random walker using an inter-module codebook that contains module names and intra-module codebooks that contain node names and an ``exit'' keyword. The codewords to the left and right of the arrow for each module respectively show the corresponding module's name in the inter-module codebook and its ``exit'' keyword in its intra-module codebook. (d) Introducing four discrete metadata values, depicted by shape, to the nodes of the network. Our extension of the traditional map equation with the content map equation additionally encodes the metadata value at each step of the random walk by introducing metadata codebooks. The metadata codewords are underlined, and keys for the metadata codebooks are given by the hollowed-out shapes. Because the blue module contains only one distinct metadata value, encoding the walker's entrance to and exit from the blue module fully specifies the metadata without requiring a metadata codebook. [Panels (a)--(c) are from a figure by Rosvall and Bergstrom \cite{rosvall_maps_2008}, Copyright (2008) National Academy of Sciences, U.S.A.]}
  \label{fig:example_encoding}
\end{figure*}

The traditional map equation is concerned solely with topology; only the path of the random walker must be encoded. To extend the map equation to networks annotated with metadata, \textit{we additionally require that the value of the metadata at each step of the random walk be encoded.} The game of the encoder is to identify which node a random walker is at for each step of the walk. Like in the traditional map equation, the encoder must report whenever the random walker changes modules. Additionally, the encoder must ensure that the metadata is fully specified by reporting its value at each step of the random walk in modules that contain more than one distinct metadata value. We require that the metadata values be encoded uniquely within each module, a requirement that, as we will later see, favors network partitions in which module labels align with metadata labels. Figure~\ref{fig:example_encoding} gives an example illustrating how our formulation extends the traditional map equation by fully specifying the metadata at each step of a random walk.

To model network dynamics, we consider a random surfer on the network. With probability $1 - \tau$, the surfer behaves like a random walker, choosing to walk along an outgoing edge of its current node with probability proportional to the outgoing edge weights. With probability $\tau$, the surfer teleports to an arbitrarily chosen node selected uniformly at random in the network. Although unnecessary in undirected components, teleportation guarantees desirable properties of the random walk in directed networks such as not becoming stuck at a node with no outgoing edges. Considering this surf in the limit of an infinite number of steps, we arrive at a steady state distribution $p_\alpha$ for every node $\alpha$ in the network. For notational convenience, we normalize the outgoing edge weights, $w_{\alpha\beta}$, from node $\alpha$ to node $\beta$ so that $\sum_\beta w_{\alpha \beta} = 1$ for every $\alpha$. We let $U$ denote a finite, discrete set of all metadata labels and assume that each node $\alpha$ is tagged with exactly one $u_{\alpha} \in U$.

\textit{The content map equation models entropy generated by the random surfer's movements between modules and within modules identically to the traditional map equation.} Between modules, we encode whenever the random surfer exits one module and enters another module. The chance at any given step that the surfer exits module $i$ is
$$q_{i\curvearrowright} = \tau \frac{n-n_i}{n-1} \sum\limits_{\alpha \in i}p_{\alpha} + (1 - \tau)\sum\limits_{\alpha \in i} \sum\limits_{\beta \notin i} p_{\alpha} w_{\alpha \beta},$$
where $n_i$ is the number of nodes in community $i$. We denote the total chance at any given step that the random surfer exits a module as
$$q_{\curvearrowright} = \sum_{i=1}^m q_{i \curvearrowright}.$$
By Shannon's source coding theorem, the minimum entropy to encode the transitions between modules, the encoding we call the ``inter-module codebook,'' is
$$H(\mathcal{Q}) = -\sum_{i=1}^m \frac{q_{i \curvearrowright}}{q_{\curvearrowright}} \log(\frac{q_{i \curvearrowright}}{q_{\curvearrowright}}).$$
The random surfer's movement within modules is another source of entropy in the map equation. Within each module, we encode the name of each node $\alpha$ that the random surfer visits with steady-state frequency $p_{\alpha}$, and we use a special ``exit'' keyword occurring with frequency $q_{i\curvearrowright}$ to encode when the random surfer exits the module. Together, these terms give the intra-module entropy for module $i$ weight $$p_{\circlearrowright}^i = q_{i \curvearrowright} + \sum\limits_{\alpha \in i} p_{\alpha}.$$ By Shannon's source coding theorem, the minimum entropy to encode the transitions within a module, an encoding we call an ``intra-module codebook,'' is
$$H(\mathcal{P}^i) = -\frac{q_{i \curvearrowright}}{p_{\circlearrowright}^i}\log\left(\frac{q_{i \curvearrowright}}{p_{\circlearrowright}^i}\right) - \sum\limits_{\alpha \in i} \frac{p_\alpha}{p_{\circlearrowright}^i}\log\left(\frac{p_\alpha}{p_{\circlearrowright}^i}\right).$$

\textit{The content map equation additionally models the entropy of the node metadata values at each step of the random surf.} Within each module $i$, we assign a codeword to each metadata value $u \in U$ that occurs with frequency $$r_u^i = \sum_{\substack{\alpha \in i, \\ u_{\alpha} = u}} p_{\alpha},$$ and we let the total metadata weight of module $i$ be
$$r_{\circlearrowright}^i = \sum_{u \in U} r_u^i = \sum_{\alpha \in i} p_\alpha.$$ 
By Shannon's source coding theorem, the minimum entropy to encode the metadata values within module $i$, in that module's ``metadata codebook,'' is $$H(\mathcal{R}^i) = -\sum\limits_{u \in W} \frac{r_u^i}{r_{\circlearrowright}^i}\log\left(\frac{r_u^i}{r_{\circlearrowright}^i}\right).$$ By encoding the metadata values separately within each module, we reward partitions whose module labels align with the metadata values. Under this encoding method, if all nodes in a module have the same metadata value, the module name in the inter-module codebook alone thereby fully specifies the metadata value at each within-module movement step, and the metadata codebook contributes zero additional entropy for this module.

Summing the entropies of the inter-module codebook, the intra-module codebooks, and the metadata codebooks, weighted by their frequency of use, the corresponding content map equation for a given partition M becomes
\begin{equation}
\label{eq:MME}
L(\text{M}) = q_{\curvearrowright} H(\mathcal{Q}) + \sum\limits_{i=1}^m p_{\circlearrowright}^i H(\mathcal{P}^i) + \eta \sum\limits_{i=1}^m r_{\circlearrowright}^i H(\mathcal{R}^i)\,
\end{equation}
where we introduce the parameter $\eta$ to control the relative weight of the metadata entropy. By increasing $\eta$, we increasingly favor communities of nodes with shared metadata values. The special case $\eta=1$ is identical to the method proposed by Smith \textit{et al}. \cite{smith_partitioning_2016}. When each module contains only a single distinct metadata label, or when $\eta = 0$, the corresponding map equation reduces to the traditional map equation.

As one way to interpret $\eta$, consider sending our message encoding the random surf over two different discrete channels, one containing the information of the traditional map equation and the other containing the metadata information. If we suppose that there are different costs to access the two channels, we can interpret $\eta$ as the relative cost to access the discrete channel of metadata information. In this interpretation, $\eta$ is an \textit{ad hoc}, relative penalty; we are not deriving $\eta$ from the dynamics of the random surf. By itself, the metadata channel does not contain useful information because the metadata codewords are module-dependent. For all finite values of $\eta$, however, the entropy of the traditional map equation contributes to our objective function, and we can assume that the receiver has access to the channel with module information.

One can imagine other ways to extend the map equation with metadata. As one example, instead of requiring that the encoder report the metadata value at each step of the random walk, one might only require the encoder to report when the metadata value changes. Rather than penalizing entropy in the metadata composition of a module, as our framework does, such a formulation would penalize the entropy of neighboring nodes that have different metadata values. As another example, instead of partitioning the nodes of the network, one might instead partition the edges of the network. An analogous coding game could be played with attributed edges to identify hierarchical and overlapping community structure; for example, see the edge partitioning methods of Ahn \textit{et al.} \cite{Ahn2010} and Kim and Jeong \cite{PhysRevE.84.026110}. We leave the study of such alternative extensions of the map equation with metadata to future work.

Throughout the paper, we compare the similarity of partitions with the scikit-learn implementation of adjusted mutual information (AMI) \cite{scikit-learn}, using the measure proposed by Vinh \textit{et al.} \cite{vinh_information_2010}: 
$$\text{AMI} = \frac{I(X, Y) - E\{I(X, Y)\}}{\text{max}\{H(X), H(Y)\} - E\{I(X, Y)\}}.$$ 
AMI adjusts the observed mutual information, $I(X, Y)$, between partitions $X$ and $Y$ by that expected at random under a hypergeometric model, $E\{I(X, Y)\}$. Normalizing by the expectation-adjusted maximum of the partitions' individual entropies, $H(X)$ and $H(Y)$, the AMI has an expected value of $0$ under randomness and a maximum value of $1$.

We use Infomap v1.0 \cite{infomapv1}, a map equation optimization software package including Eq.~(\ref{eq:MME}), for all of our experiments. Because teleportation can blur a network's modular structure \cite{PhysRevE.85.056107}, we avoid teleportation by considering undirected networks and setting the teleportation probability $\tau = 0$.

\section{Synthetic Graph Results}

\begin{figure*}
  \includegraphics[width=\linewidth]{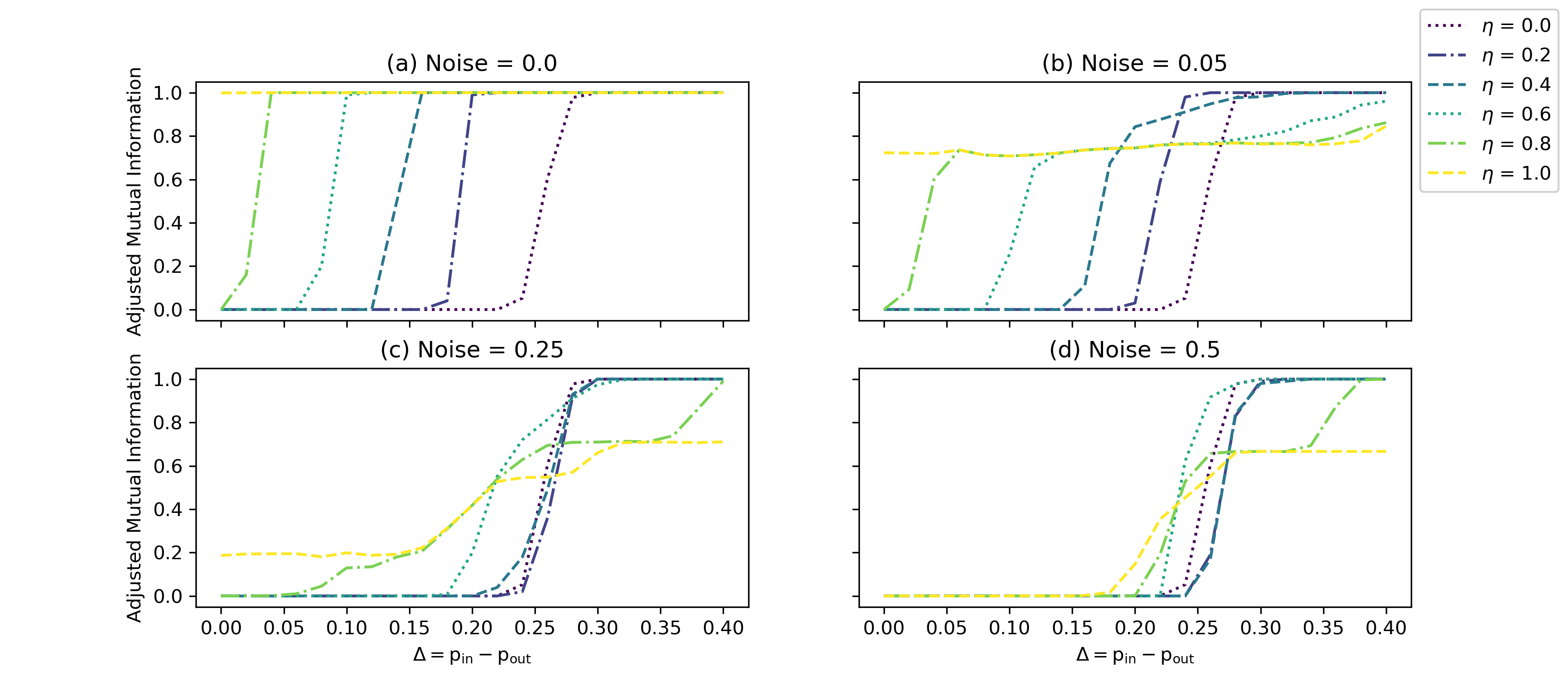}
  \caption{(Color online) Detectability experiments on a planted-partition stochastic block model with $N = 200$ nodes evenly divided into two communities. Each node's attribute label corresponds with its planted partition community with probability noise, and $\Delta$ measures the assortativity of the planted partition. We see that focusing on the metadata by increasing $\eta$ enables the algorithm to overcome the detectability limit when the metadata has strong signal, but increasing $\eta$ ceilings the algorithm's performance when the metadata is noisy.}
  \label{fig:detectability_limit}
\end{figure*}

To analyze how varying $\eta$ impacts the content map equation's ability to detect communities, we construct synthetic graphs according to a two-block planted-partition stochastic block model (SBM) with $N = 200$ nodes evenly divided into $2$ communities, where an edge connecting two nodes in the same community exists with probability $p_{in}$, and an edge connecting two nodes in different communities exists with probability $p_{out}$. We additionally annotate each node with one of two discrete attribute labels based on the ``noise'' parameter. Each node with probability $1 - \mathrm{noise}$ is assigned an attribute label equating to its community assignment and with probability noise is assigned the opposite attribute label. With $\mathrm{noise}=0$, the communities and attributes correspond perfectly. With $\mathrm{noise} = 0.5$, the attributes are totally random. Figure~\ref{fig:detectability_limit} shows results for different $\eta$ exploring the AMI between the planted partition and the partition identified by optimizing the content map equation, where each data point is the average of 100 trials with edge density $\rho = (p_{in} + p_{out})/2 = 0.2$. We show a maximum value of $\eta = 1$ because it is the same as the results for higher values of $\eta$.

For the corresponding unannotated SBM, it has been shown in the limit as $N \to \infty$ that the two-block planted-partition structure becomes undetectable for $\Delta = p_{in} - p_{out}$ below the threshold given by $N\Delta^* = \sqrt{4N\rho(1-\rho)}$. (For more detail, see \cite{decelle_2011,nadakuditi_2012} and the discussion including non-sparse and multilayer networks in \cite{taylor_2016}.) Communities are only detectable when $\Delta > \Delta^*$ because otherwise the community structure is too weak relative to the background noise of the generative model.
For the parameters of our experiment, $\Delta^* \doteq 0.057$. The detectability of partitions, however, is distinct from resolution selection, i.e., determining the size of partitions. Experimentally, we find that Infomap, partitioning solely by network structure with $\eta = 0$, transitions from returning a single-community partition at $\Delta = 0.22$ to returning the planted two-community partition at $\Delta = 0.3$. For the remainder of this discussion, we refer to the $\Delta = [0.22, 0.3]$ region as the ``selection threshold''.

In the presence of metadata signal below the selection threshold, we find as expected that increasing $\eta$ increases AMI. Although the communities are undetectable (or at least not selected) based on the network connectivity alone, the metadata provides additional information. Moreover, as the metadata becomes more aligned with the communities, it provides a greater boost to the algorithm's performance. For example, as Fig.~\ref{fig:detectability_limit} illustrates, reducing the noise from 0.25 to 0.05 increases the average AMI at $\eta = 1$ from around $0.2$ to around $0.7$.

Our experiments show both that increasing $\eta$ can benefit AMI when the community structure has relatively low community assortativity, i.e., when $\Delta$ is small, and that increasing $\eta$ can hurt AMI when the communities have relatively high assortativity, i.e., when $\Delta$ is large. When $\Delta$ is small, increasing $\eta$ allows the algorithm to detect the signal present in the metadata, which is greater than that present in the network structure. But when $\Delta$ is large, increasing $\eta$ too much causes the algorithm to overfit the metadata and miss the communities present in the network structure. This effect can be seen in Fig.~\ref{fig:detectability_limit} at $\mathrm{noise}=0.25$, where below the selection threshold high values of $\eta$ achieve average AMI of $0.2$ compared to average AMI of $0$ for low values of $\eta$, while above the selection threshold the highest value of $\eta$ has average AMI capped at $0.7$ while lower values of $\eta$ achieve average AMI approaching the perfect score, $1$.

Perhaps surprisingly, increasing $\eta$ increases AMI below the selection threshold even when the metadata is totally random, i.e., when $\mathrm{noise} = 0.5$. In this case, increasing $\eta$ appears to act as an effective resolution parameter. By encouraging partitions with more homogeneous metadata values, increasing $\eta$ makes the algorithm prefer a partition containing several smaller communities over the larger, single-community partition selected without using metadata. As a result, increasing $\eta$ when below the selection threshold moves the algorithm away from the single-community partition, which has an AMI of $0$, to a partition of several communities with positive AMI.

\section{Real-World Graph Results}

\subsection{Lazega Lawyers Networks of Law Firm Relationships}

The Lazega lawyers networks consist of 71 lawyers at a corporate lawfirm in the American Northeast \cite{lazega_collegial_2001}. Surveys were conducted to form the basis of three networks connecting the same people: the coworking network, based on a survey question asking each lawyer with whom in the firm the lawyer has worked; the advice network, based on a survey question asking each lawyer to whom in the firm the lawyer has gone for professional advice; and the social network, based on a survey question asking each lawyer with whom in the firm the lawyer socializes outside of work. As node metadata, we additionally use information that each lawyer reported about the lawyer's status (partner or associate), gender (man or woman), office (Boston, Hartford, or Providence), practice (litigation or corporate), and law school (Harvard / Yale, University of Connecticut, or other). 

\begin{figure*}
  \includegraphics[width=\linewidth]{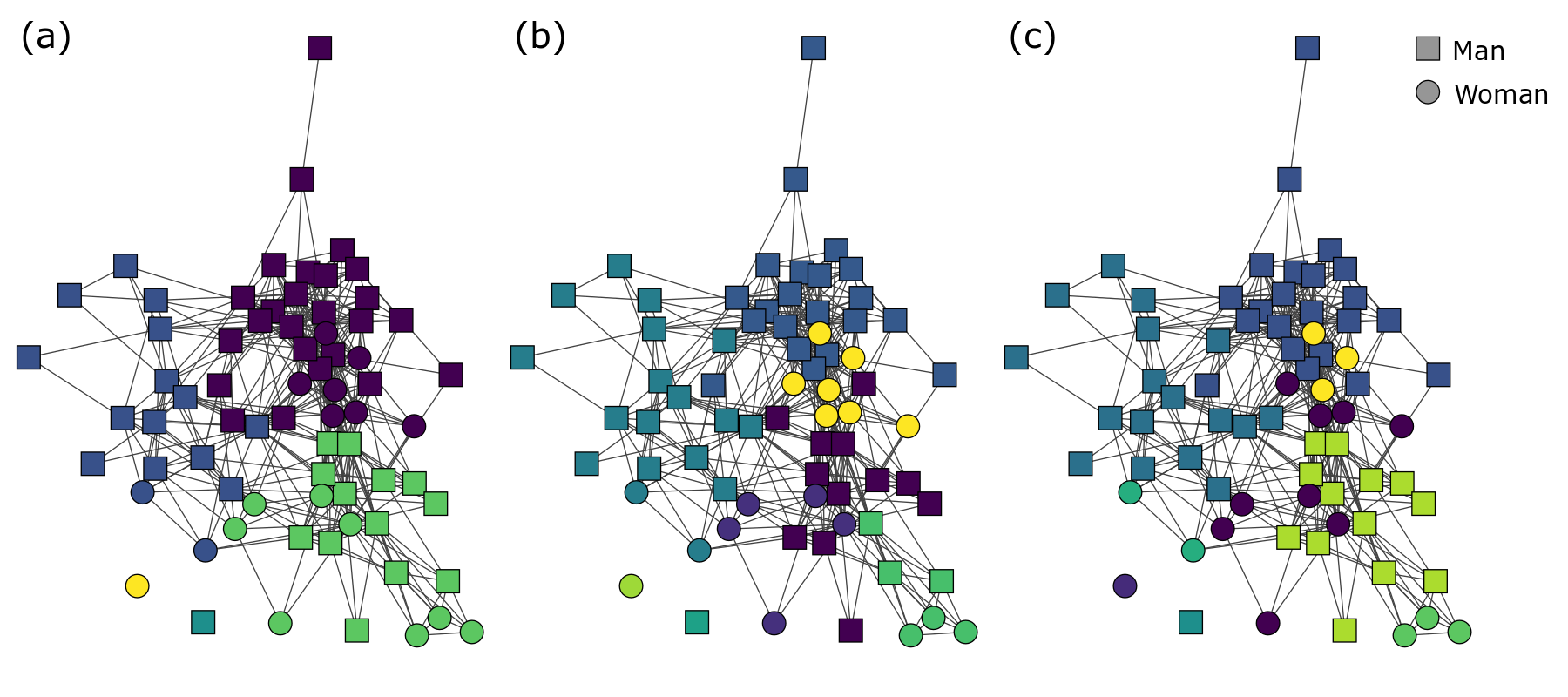}
  \caption{(Color online) The Lazega lawyers friendship network partitioned with the metadata attribute gender at (a) $\eta = 0$, (b) $\eta = 0.7$, and (c) $\eta = 1.25$. Color encodes the partition while shape encodes the metadata.}
  \label{fig:eta_lens}
\end{figure*}

Figure \ref{fig:eta_lens} illustrates how increasing $\eta$ affects the returned network partition. The figure shows communities in the friendship network using the metadata attribute gender. Node shapes encode metadata values while node colors encode the algorithm's partition. At $\eta = 0$, the algorithm optimizes for the traditional map equation, returning a partition based solely on network topology. With increasing $\eta$, the algorithm returns modules more aligned with the metadata. For example, moving from $\eta = 0$ to $\eta = 0.7$, two modules of all men and two modules of all women emerge, but two modules remain containing both men and women. At $\eta = 1.25$, the modules are either all-man or all-woman, and the metadata codebooks contribute zero additional entropy to the content map equation. Note, however, that even as the algorithm increasingly takes the metadata into consideration with increasing $\eta$, the algorithm still respects the topology of the network because the random walker proceeds independently of node metadata values.

\begin{figure*}
  \includegraphics[width=\linewidth]{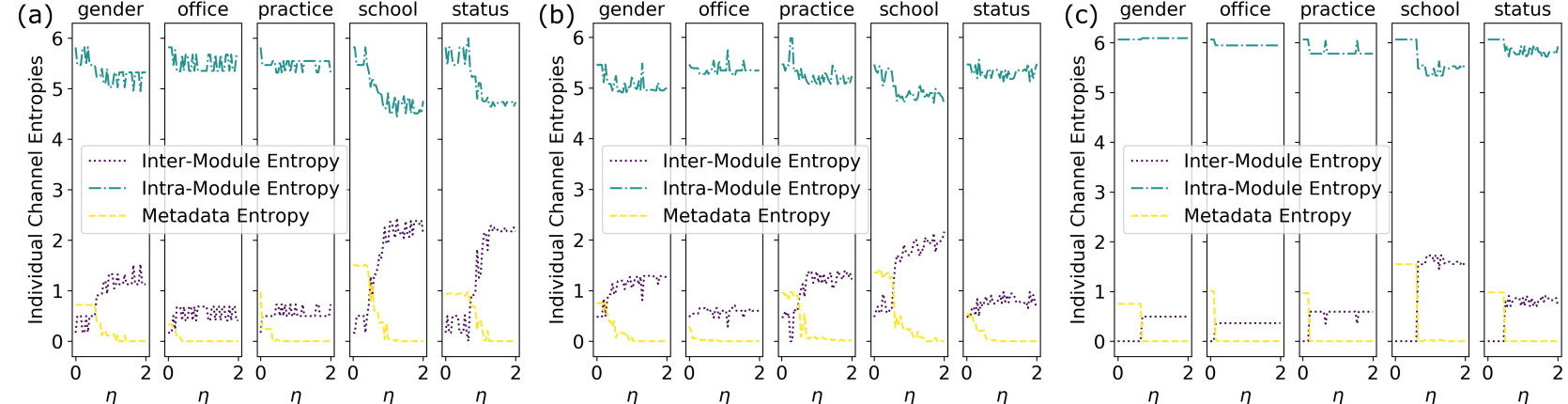}
  \caption{(Color online) Sums of various types of entropy when partitioning the Lazega lawyers (a) coworking, (b) friendship, and (c) advice networks. The sums are weighted by frequency of codebook use but not weighted by $\eta$.}
  \label{fig:lazega_channel_entropies}
\end{figure*}

Figure~\ref{fig:lazega_channel_entropies} shows the sum of the entropies of each codebook type, weighted by frequency of use but not relatively weighted by $\eta$, for partitions of the Lazega lawyers networks at varying $\eta$ for the different metadata types. ``Inter-module entropy'' measures the first term of Equation \ref{eq:MME}, ``intra-module entropy'' measures the second term of Equation \ref{eq:MME}, and ``metadata entropy'' measures the third term of Equation \ref{eq:MME}, unweighted by $\eta$. In all the plots, metadata entropy is at its maximum when $\eta = 0$ and decreases until the metadata entropy becomes $0$ for sufficiently large $\eta$. With perfect optimization of the content map equation, increasing $\eta$ would cause a strict decrease in the metadata entropy. However, the stochastic, approximate optimization algorithm we employ causes Fig.~\ref{fig:lazega_channel_entropies} to deviate from strict monotonicity in the metadata entropy.

As Fig.~\ref{fig:lazega_channel_entropies} illustrates, once the relative weight of the metadata codebook is sufficiently large, an optimal partition's metadata codebook will necessarily have zero entropy. Optimizing the content map equation in the limit as $\eta \to \infty$ becomes a constrained optimization of the traditional map equation. A candidate optimal partition must have only one metadata attribute per module, and the optimal partition is the partition from this constrained region of partition space optimizing the traditional map equation.

The Lazega lawyers networks, which share the same set of attributed nodes but have different edge types, allow us empirically to study how different edge formation processes influence metadata community detection. In Fig.~\ref{fig:lazega_channel_entropies}b and Fig.~\ref{fig:lazega_channel_entropies}c, the gender panel shows how different connectivity patterns among the same set of attributed nodes can lead to qualitatively different behavior with increasing $\eta$. In the Fig.~\ref{fig:lazega_channel_entropies}c advice network, increasing $\eta$ results in a sharp transition from the topological partition at $\eta = 0$ to the partition with zero metadata entropy that is optimal as $\eta \to \infty$; the algorithm finds no intermediate partitions. In the Fig.~\ref{fig:lazega_channel_entropies}b friendship network, on the other hand, the transition is more gradual. As the algorithm transitions from $\eta = 0$ to the limit as $\eta \to \infty$, it returns multiple intermediate partitions such as the one shown in Fig.~\ref{fig:eta_lens}b. Furthermore, Fig.~\ref{fig:lazega_channel_entropies} suggests that topological network properties govern whether the metadata entropy decrease is gradual or sharp. All of the transitions for the Fig.~\ref{fig:lazega_channel_entropies}a coworking and Fig.~\ref{fig:lazega_channel_entropies}b friendship networks are gradual, whereas all of the transitions for the Fig.~\ref{fig:lazega_channel_entropies}c advice network are sharp.

\begin{figure*}
  \includegraphics[width=\linewidth]{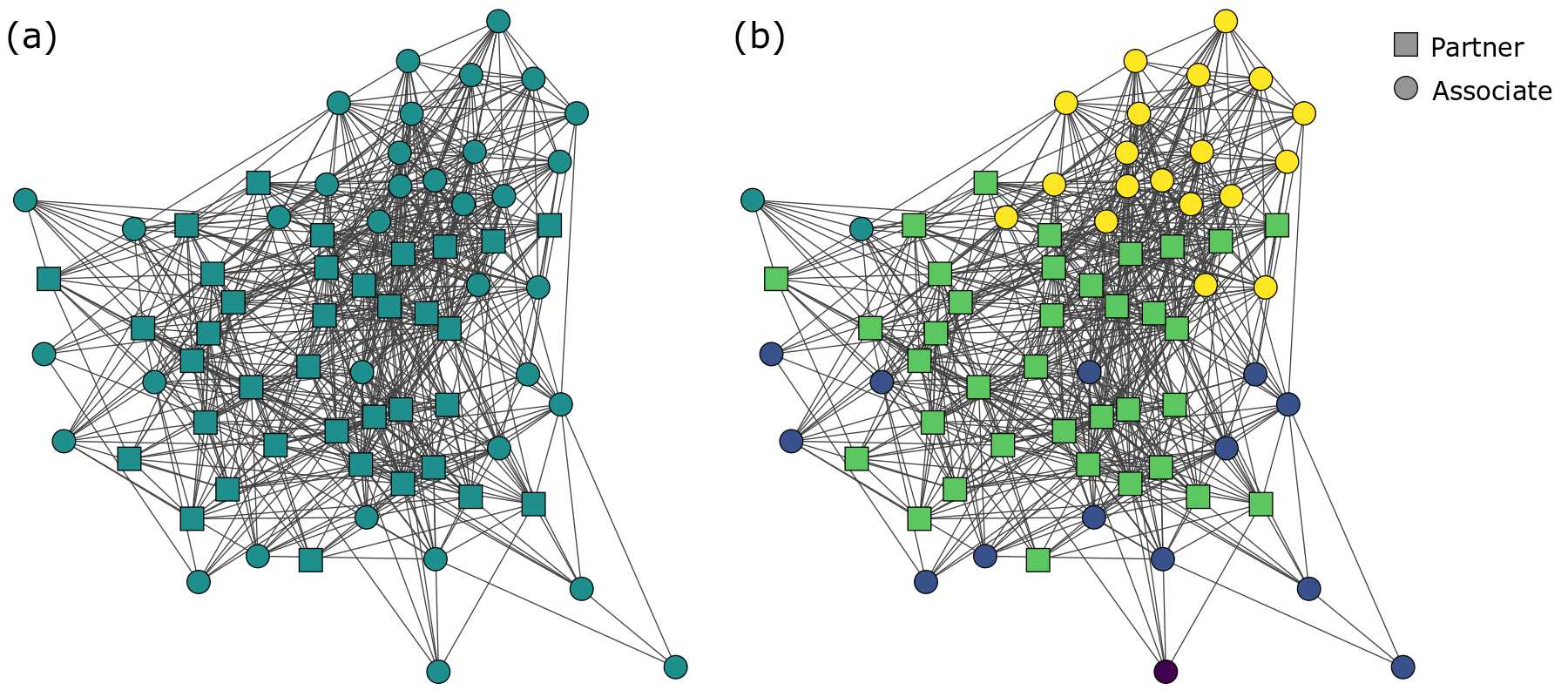}
  \caption{(Color online) The Lazega lawyers advice network partitioned with the metadata attribute status at (a) $\eta = 0$ and (b) $\eta = 0.5$. Color encodes the partition while shape encodes the metadata.}
  \label{fig:advice_by_status}
\end{figure*}

Each panel of Fig.~\ref{fig:lazega_channel_entropies} summarizes the entropies of an entire set of partitions that can be studied in more depth. For example, consider Fig.~\ref{fig:advice_by_status}, which shows partitioning the Lazega lawyers advice network with metadata about each lawyer's status in the firm, either partner or associate. Partitioning only on network topology in Fig.~\ref{fig:advice_by_status}a at $\eta = 0$, the algorithm returns a partition with only one module. But in Fig.~\ref{fig:advice_by_status}b at $\eta = 0.5$, we see the best partition of the network that puts partners and associates into separate modules; while there is one module of partners, there are four modules of associates. In other words, when we constrain the map equation to modules of all partners and all associates, the best description of the flow of advice in the network has one module of partners and four modules of associates.

Perhaps the partners of the firm, who have presumably been around the longest, have spent enough time together that each partner trusts the other partners for professional advice, whereas the associates of the firm have not yet developed trust with all the other associates. Or perhaps the partners of the firm are the most knowledgeable about the firm's operations and form a core module of nodes in the advice network with the associate modules at the periphery. Whatever the cause of the difference between the number of partner and associate modules, this difference is an interesting structure in the network that is highlighted by partitioning with node metadata, motivating potential follow-up study.

\subsection{Add Health Network of High School Friendship}

The high school friendship network used here is from the US National Longitudinal Study of Adolescent Health and was provided by the Add Health project of the Carolina Population Center.
Each of the 795 nodes of the graph is a student in an American middle school (7-8th grade, 12-14 years of age) and corresponding high school (9-12th grade, 14-18 years of age). Edges between nodes represent friendships determined by survey. As metadata for each node, we use the student survey data of grade (range 7-12), race (``white only'', ``black only'', ``any Hispanic'', ``Asian only,'' or ``mixed / other''), school code (middle or high school), and sex (male or female).

The presence of various metadata types allows us to highlight a key feature of the algorithm, that it allows tuning $\eta$ to see how the network partitions under a particular metadata type of interest. In prior work on community detection with metadata, the method of Newman and Clauset \cite{newman_structure_2016} was applied to the network three times, separately using grade, race, and sex, in each case partitioning the network into two communities. Using grade metadata, the algorithm splits the network into clear middle school and high school groups. Similarly, the algorithm divides the network into a predominantly white and a predominantly black group when it uses race metadata. However, when asked to use sex metadata, the algorithm of Newman and Clauset ignores the sex metadata because the metadata does not have a strong enough correlation with the network structure. As Newman and Clauset note, for someone interested only in the metadata to the extent that it correlates with network structure, it is advantageous for the algorithm to disregard metadata that does not correlate.

But suppose that \textit{a priori} a network analyst knows she cares about structures related to a particular metadata type. In our example, a social science researcher might be interested in how the high school friendship network organizes by sex, however strong or weak the sex partition might be. Using the algorithm of Newman and Clauset, there is no way for such a researcher to convey to the algorithm this preference for the sex metadata type. A key feature of our metadata map equation is the ability, using $\eta$, to specify the weights of given metadata types relative to the network topology in assigning communities.

\begin{figure*}
  \includegraphics[width=\linewidth]{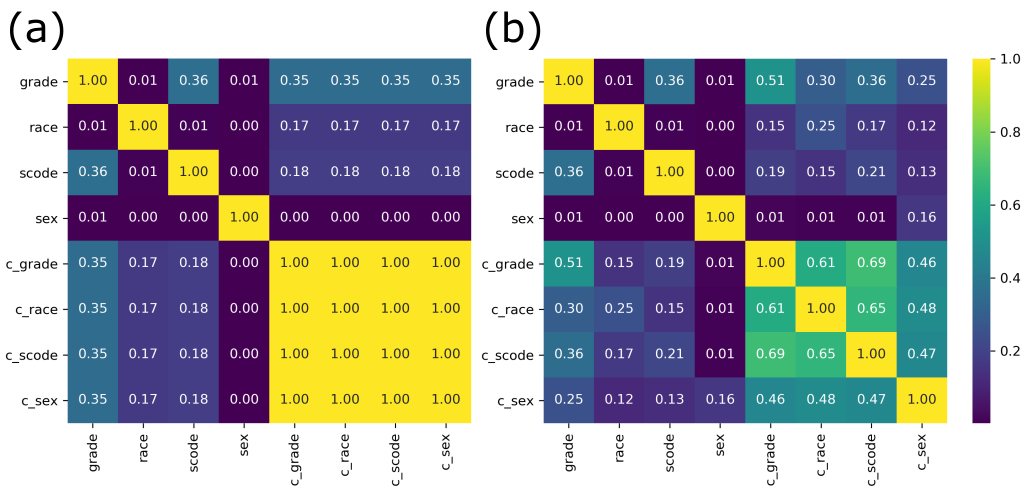}
  \caption{(Color online) Pairwise AMIs of High School Social Network partitions at (a) $\eta = 0$ and (b) $\eta = 3$. For example, ``grade'' is the partition given by each node's grade, and ``c\_grade'' is the algorithm's returned partition when partitioning with the grade metadadta.}
  \label{fig:addhealth_pairwise_mats}
\end{figure*}

Figure \ref{fig:addhealth_pairwise_mats} demonstrates how our algorithm can specify a relative weighting for various metadata types. When $\eta = 0$, all of the partitions follow only the network topology. In that case, our results, consistent with those of the algorithm of Newman and Clauset, show that the metadata attributes of grade, school code, and race have the highest mutual information with the topological partition, with respective AMI values of $0.35$, $0.18$, and $0.17$, while the metadata attribute of sex has the least mutual information with the topological partition, with an AMI of $0$. When we increase $\eta$ to $\eta = 3$, we see using each of the metadata values (grade, race, school code, sex) that the algorithm finds a partition of the network that, compared to the community detection done with only the network topology at $\eta = 0$, has increased AMI with the metadata.

Importantly, the partitioning of the high school network with a relative metadata channel weight of $\eta = 3$ does not simply ignore the network structure. Consistent with the results of Newman and Clauset, we see that grade is the metadata value for which we can achieve the highest AMI between the algorithm's partition and the node metadata, with an AMI of $0.51$, and we find that our algorithm's partition using sex has the least correspondence with the node metadata, an AMI of $0.16$.

\begin{figure}
  \includegraphics[width=\linewidth]{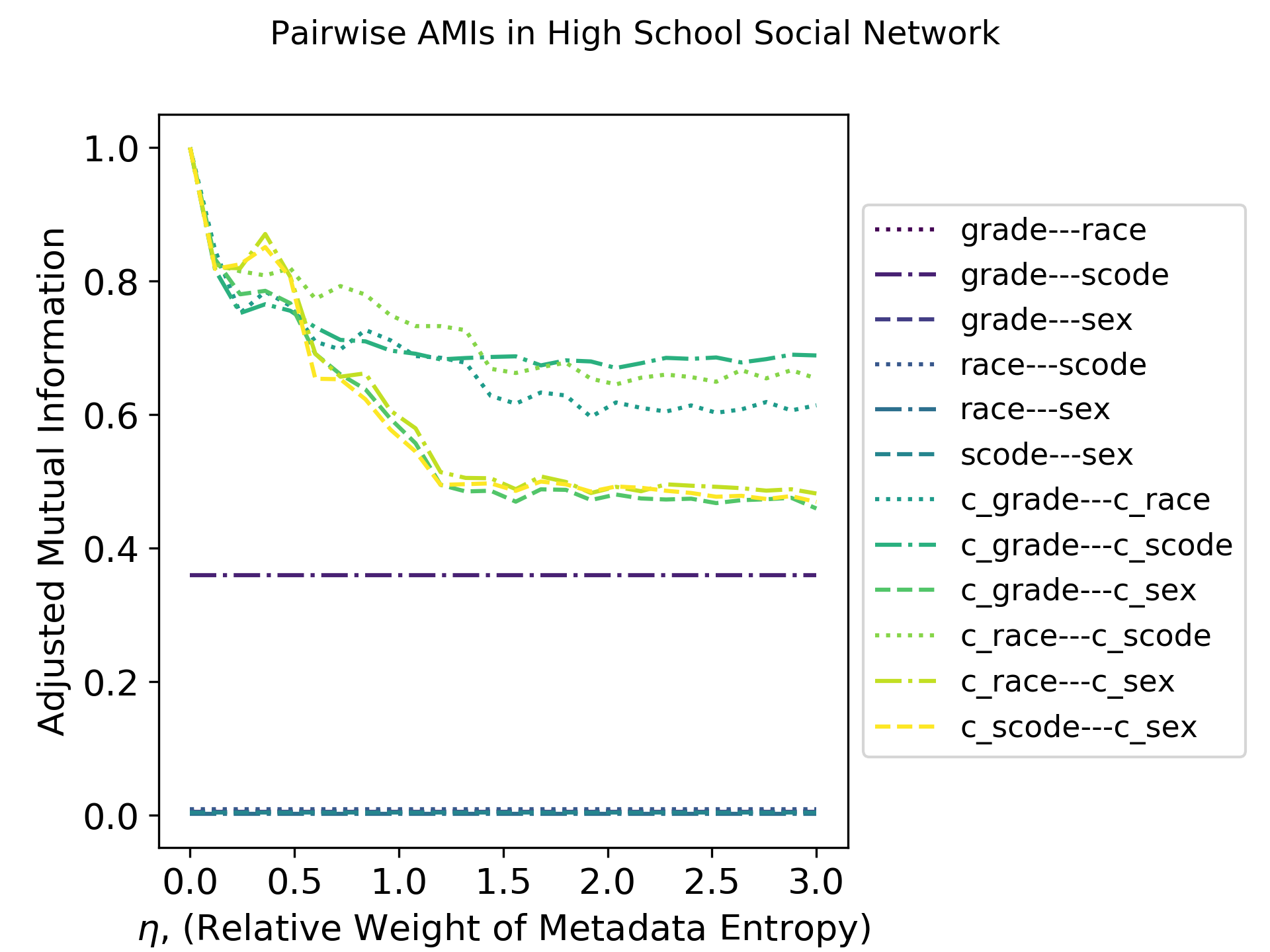}
  \caption{(Color online) Pairwise AMIs of high school social network partitions. For example, ``grade'' is the partition given by each node's grade, and ``c\_grade'' is the algorithm's returned partition when partitioning with the grade metadata for a given value of $\eta$.}
  \label{fig:addhealth_pairwise_c_amis}
\end{figure}

Figure \ref{fig:addhealth_pairwise_c_amis} illustrates the role of $\eta$ and the metadata in the community detection process. Each point on the graph of Fig.~\ref{fig:addhealth_pairwise_c_amis} is an AMI calculation. ``Grade'', ``race,'' ``scode'' (school code), and ``sex'' are the partitions of the network given by the respective metadata labels, and ``c\_grade,'' ``c\_race,'' ``c\_scode,'' and ``c\_sex'' are partitions returned by the algorithm given the corresponding metadata type as input and the value of $\eta$ indicated by the x-axis. The lines of Fig.~\ref{fig:addhealth_pairwise_c_amis} show how the AMIs of pairs of these partitions change with $\eta$. Pairs of partitions determined solely by metadata are constant with respect to $\eta$ because the metadata of each node is fixed. Pairs of community detection partitions considering different metadata begin with an AMI of 1 because, at $\eta = 0$, the algorithm only considers network topology. As $\eta$ increases, the algorithm returns partitions more aligned with the attribute under consideration, and the pairwise AMIs of these partitions decrease.

One can suppose that the optimal partition at $\eta = 0$ is a point in the space of all possible partitions of the graph. In this interpretation, increasing $\eta$ for a given metadata type causes the optimal partition to shift in partition space toward partitions more aligned with the particular metadata type. As the optimal partitions for different metadata types undergo such shifts, they diverge in partition space, and as Fig.~\ref{fig:addhealth_pairwise_c_amis} illustrates, their pairwise AMI decreases.

\begin{figure}
  \includegraphics[width=\linewidth]{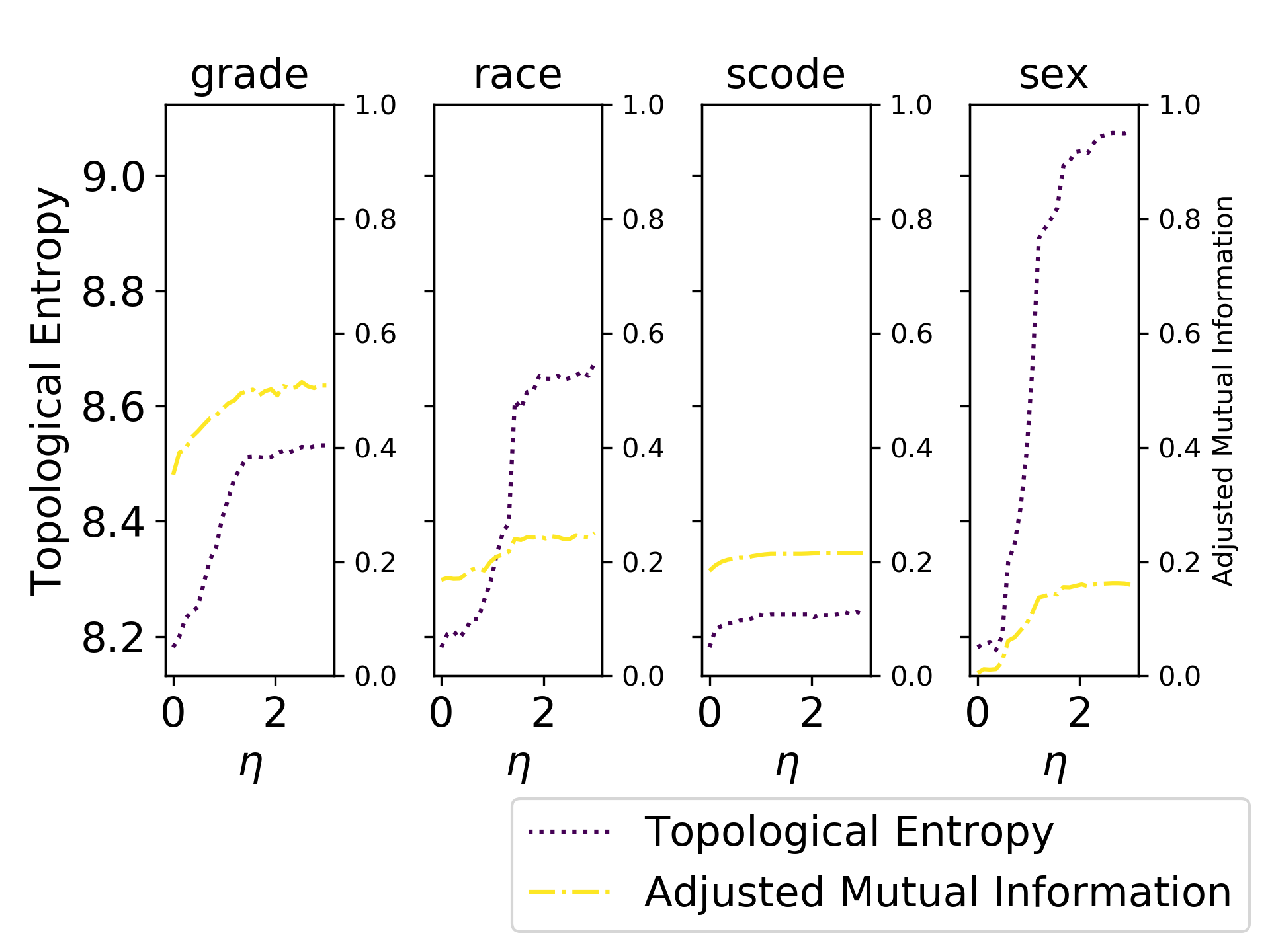}
  \caption{(Color online) Topological entropy and AMI tradeoff when partitioning with metadata in the high school social network. Topological entropy, equal to the traditional map equation, is the sum of the inter-module codebook and intra-module codebook entropies. The AMI is between the node metadata and the algorithm's returned partition.}
  \label{fig:addhealth_tradeoff}
\end{figure}

Fig.~\ref{fig:addhealth_tradeoff} shows another way to understand $\eta$. One can consider increasing $\eta$ as paying a topological entropy price (the sum of the inter-module codebook and intra-module codebook entropies, which is equal to the traditional map equation) for increased AMI with the metadata. The varying shapes of the curves in Fig.~\ref{fig:addhealth_tradeoff} show how the price of this tradeoff at a given value of $\eta$ depends on how node metadata values relate to the network structure. For example, consider the curves corresponding to the school code and sex metadata. For school code metadata, the optimal partition at $\eta = 0$ is already relatively close to meeting the constraint required as $\eta \to \infty$ that each module have just one metadata attribute. One cannot trade topological entropy for much increase in AMI with the metadata because increasing $\eta$ does not much change the returned partition. For sex metadata, however, the optimal partition at $\eta = 0$ is relatively far from having just one metadata attribute per module. By increasing $\eta$, one can pay topological entropy for increased AMI with the metadata as the returned partition shifts toward obeying the constraint imposed as $\eta \to \infty$.

\section{Conclusion}

We introduced a tuning parameter to the content map equation that explicitly controls the importance of metadata relative to edge connectivity in community detection. We demonstrated on synthetic graphs how focusing on the metadata can overcome the detectability limit when the metadata is well-aligned with the topological community structure and also how focusing on the metadata can put a ceiling on performance when the metadata is misaligned with the topological community structure. On real-world graphs, we demonstrated how a practitioner might tune the content map equation to ``zoom in'' and ``zoom out'' on communities with varying levels of metadata focus.

Our method probes the relationship between community structure and metadata. While we gave the algorithm only one type of metadata attribute at a time, future work might simultaneously incorporate metadata attributes of different types and relative weightings. It might also be interesting to study how various metadata types relate to various network processes as, for example, different metadata types might relate to the spread of different kinds of information.

\section*{Acknowledgements}

We are grateful to Zachary Boyd, Peter Diao, Ryan Gibson, and William Weir for discussions throughout the creation of this work. We thank Daniel Edler, Martin Rosvall, and the entire team at the Integrated Science Lab of Ume\aa\ University for their collaboration incorporating our work into the Infomap software package. Research reported in this publication was supported by the James S. McDonnell Foundation 21st Century Science Initiative - Complex Systems Scholar Award grant \#220020315. The content is solely the responsibility of the authors and does not represent the official views of the sponsor.

This research uses data from Add Health, a program project directed by Kathleen Mullan Harris and designed by J. Richard Udry, Peter S. Bearman, and Kathleen Mullan Harris at the University of North Carolina at Chapel Hill, and funded by grant P01-HD31921 from the Eunice Kennedy Shriver National Institute of Child Health and Human Development, with cooperative funding from 23 other federal agencies and foundations. Special acknowledgment is due Ronald R. Rindfuss and Barbara Entwisle for assistance in the original design. Information on how to obtain the Add Health data files is available on the Add Health website (http://www.cpc.unc.edu/addhealth). No direct support was received from grant P01-HD31921 for this analysis.

\bibliography{references}

\end{document}